\documentclass[runningheads,a4paper]{llncs}

\usepackage{amssymb}
\usepackage{graphicx}
\usepackage{multirow}
\usepackage{tabu}
\usepackage[table]{xcolor}
\usepackage{amsmath}
\usepackage{program}
\usepackage{multirow,array}
\usepackage{flushend}
\usepackage{enumitem}

\graphicspath{{./pic_wei/}}

\usepackage{array}
\newcolumntype{L}[1]{>{\raggedright\let\newline\\\arraybackslash\hspace{0pt}}m{#1}}
\newcolumntype{C}[1]{>{\centering\let\newline\\\arraybackslash\hspace{0pt}}m{#1}}
\newcolumntype{R}[1]{>{\raggedleft\let\newline\\\arraybackslash\hspace{0pt}}m{#1}}

\begin{document}

\mainmatter 

\title{Multiple Fault Attack on PRESENT with a Hardware Trojan Implementation in
FPGA}
\titlerunning{Multiple Fault Attack on PRESENT}

\author{Jakub Breier and Wei He}
\institute{Physical Analysis and Cryptographic Engineering, \\
Temasek Laboratories at Nanyang Technological University\\
Singapore\\
\{jbreier, he.wei\}@ntu.edu.sg}
\maketitle

\begin{abstract}
Internet of Things connects lots of small constrained devices to the Internet.
As in any other environment, communication security is important and
cryptographic algorithms are one of many elements that we use in order to keep
messages secure.
Because of the constrained nature of these environments, it is necessary to use algorithms that do not require high computational power.
Lightweight ciphers are therefore ideal candidates for this purpose. 

In this paper, we explore a possibility of attacking an ultra-lightweight cipher PRESENT
by using a multiple fault attack. Utilizing the Differential Fault Analysis technique, we were able to
recover the secret key with two faulty encryptions and an exhaustive search of $2^{16}$ remaining key bits.
Our attack aims at four nibbles in the penultimate round of the cipher, causing faulty output in all nibbles of
the output. We also provide a practical attack scenario by exploiting Hardware Trojan (HT) technique for 
the proposed fault injection in a Xilinx Spartan-6 FPGA.  
\end{abstract}

\section{Introduction}
Internet of Things brings new challenges into the security field. With
interconnection of a huge number of small devices with constrained computational
capabilities, there is a need to design algorithms and protocols simple enough
to be run on such devices within a reasonable time frame yet still preserving
high level of security. Lightweight cryptography provides algorithms that use
operations fulfilling such requirements. It delivers adequate security and does
not always lower the security-efficiency ratio~\cite{katagi2008iot}. Currently
there are many lightweight cryptography algorithms
available, providing various security levels and
encryption speed~\cite{dinu2015lightweight}. For the further work we have chosen
the PRESENT algorithm.

PRESENT is an ultra-lightweight block cipher, introduced by Bogdanov et al. in
2007 \cite{present}. The algorithm was standardized in 2011, by the ISO/IEC
29192-2:2011 standard. It is based on SP-network, therefore uses three
operations in each round -- 4-bit non-linear substitution, bit permutation and
\texttt{xor} with the key. The best cryptanalysis presented so far is a
truncated differential attack on 26 out of 31 rounds
presented in 2014~\cite{blondeau2014present}.

Fault attacks exploit a possibility to change the intermediate values in the
algorithm execution so that it can radically reduce the key search space or even
reveal the key. The first attack was proposed by Boneh, DeMillo and Lipton in
1996 \cite{boneh_demillo}, following by the practical attack by Biham and Shamir
one year later \cite{biham_shamir}.

Currently, fault analysis is a popular method to attack cryptographic implementations, utilizing clock/voltage glitch techniques, 
diode laser and ION beam irradiation, EM pulse, or hardware trojans. For attacking symmetric block ciphers, the most popular 
technique is Differential Fault Analysis (DFA), in which the fault is usually inserted in the last rounds of a cipher for observing differences 
between correct and faulty ciphertexts. Other techniques include Collision Fault Analysis (CFA), Ineffective Fault Analysis (IFA), Safe-Error Analysis (SEA) \cite{clavier2012,danger2013}. We have chosen DFA as our attack technique, with inserting multiple faults in the penultimate round of PRESENT cipher.

In our work we present a novel multiple fault attack on PRESENT. By injecting four nibble-switch faults in the penultimate round we were able to
recover the secret key with only two faulty ciphertexts and an exhaustive search of $2^{16}$ remaining bits of the key. In both faulty ciphertexts, it 
is necessary to flip different nibbles in order to produce a different fault mask in the last round. We provide a practical attack scenario by exploiting Hardware Trojan (HT) technique for fault injection in a Xilinx Spartan-6 FPGA.

The rest of the paper is organized as follows. Section \ref{related} provides an overview of known fault attacks on PRESENT cipher and Section \ref{present}
describes this cipher in details. Our attack model is proposed in Section \ref{attackmodel} and HT implementation of our attack is described in Section
\ref{implementation}. Finally, Section \ref{conclusions} concludes our work and provides motivation for further research.

\section{Related Work}
\label{related}
The first DFA attack on PRESENT was published in 2010 by G. Wang and S. Wang
\cite{present_dfa1}. Their attack aimed at a single nibble and they were able to
recover the secret key with the computational complexity of $2^{29}$ by using 64
pairs of correct and faulty cipher texts on average.

Zhao et al.~\cite{present_dfa2} proposed a fault-propagation pattern based DFA
and demonstrated this technique on PRESENT and PRINT ciphers. The attack on
PRESENT-80 and PRESENT-128 uses 8 and 16 faulty cipher texts on average, respectively, and reduces the
master key search space to $2^{14.7}$ and $2^{21.1}$.

Gu et al.~\cite{present_dfa4} used a combination of differential fault analysis
and statistical cryptanalysis techniques to attack lightweight block ciphers.
They tested their methodology on PRESENT-80 and PRINT-48. The attack on PRESENT
is aimed at middle rounds of algorithm, using single random S-box and multiple
S-boxes fault attack. The main outcome of the paper is an extension of the fault
model from the usual fault model, which aims at ultimate or penultimate rounds,
to the other rounds as well. 

Bagheri et al.~\cite{present_dfa3} presented two DFA attacks on PRESENT, the
first one attacks a single bit and the second one attacks a single nibble of the
intermediate state. They were able to recover the secret key with 18 faulty
cipher texts on average, using the second attack.

The most efficient attack so far was proposed by Jeong et al.~\cite{present_dfa5}. They 
used a 2-byte random fault model, attacking the
algorithm state after round 28. For PRESENT-80, they needed two 2-byte faults
and for PRESENT-128 they needed three 2-byte faults and an exhaustive search of
$2^{22.3}$ on average.

Hardware Trojans have drawn much attention during the past decade due to their
severity in security-sensitive embedded systems~\cite{defense2005defense}. As an
enormous network of diverse embedded devices, Internet of Things (IoT) is populated 
with a great number of ICs, to collect, encrypt, transmit, and store data. For each 
node inside an IoT system, a complete functional chip normally consists of sorts of IPs, 
and they are typically designed and manufactured by off-shore design houses or foundries. 
In theory, any parties involving into the design or manufacturing stages can make 
alterations in the circuits for malicious purposes~\cite{king2008designing}~\cite{shiyanovskii2010process}. 
These tiny changes or extra logic can hide inside the system during the majority of its lifetime 
until a specific activation mechanism is awaken for pilfering secrets or impairing the main functionality.
  
As a typical stealthy modification to ICs, a Hardware Trojan (either named Trojan Horse) was intentionally
integrated into embedded devices for disabling or destroying a system, leaking 
confidential information from side channels, or triggering critical faults~\cite{tehranipoor2010survey}~\cite{lin2009trojan}~\cite{jin2009experiences}. 
HT can be implanted into the circuit at multiple stages with a stealthy nature. The post-manufacturing 
testing often fail to detect it since Trojan only influences the circuit under specific conditions~\cite{chakraborty2013hardware}. At a proper future 
time, the Trojan can be activated. Unlikely to the counterpart Software Trojan (ST), HT cannot be removed by upgrading the software in each device. So HT is truly furtive and ineradicable which hence poses more serious threat to the system security, particularly to the cryptographic blocks inside the IoT system. HT basically consists 
of two components: (a) the trigger signal that performs to activate the 
inserted trojan, and (b) the payload that is affected by the trojan~\cite{tehranipoor2011introduction}. Actually, many solutions have been proposed for detecting the implanted trojans, such as the fine-grained optical inspections~\cite{torrance2011state} and the \emph{side-channel} based comparison with the fully trustworthy ``golden chip''. To increase the difficulties of HT detection, the trojan size is preferably to be as small as possible, \emph{w.r.t} its host design. In this 
paper, a compact Trojan module is presented which serves to inject multiple faults into specific algorithmic points in PRESENT cipher, which makes the proposed multiple fault attack approach realistic in practices.

\section{Overview of PRESENT Cipher}
\label{present}
\begin{figure}[h]
\centering
\includegraphics[width=0.3\textwidth]{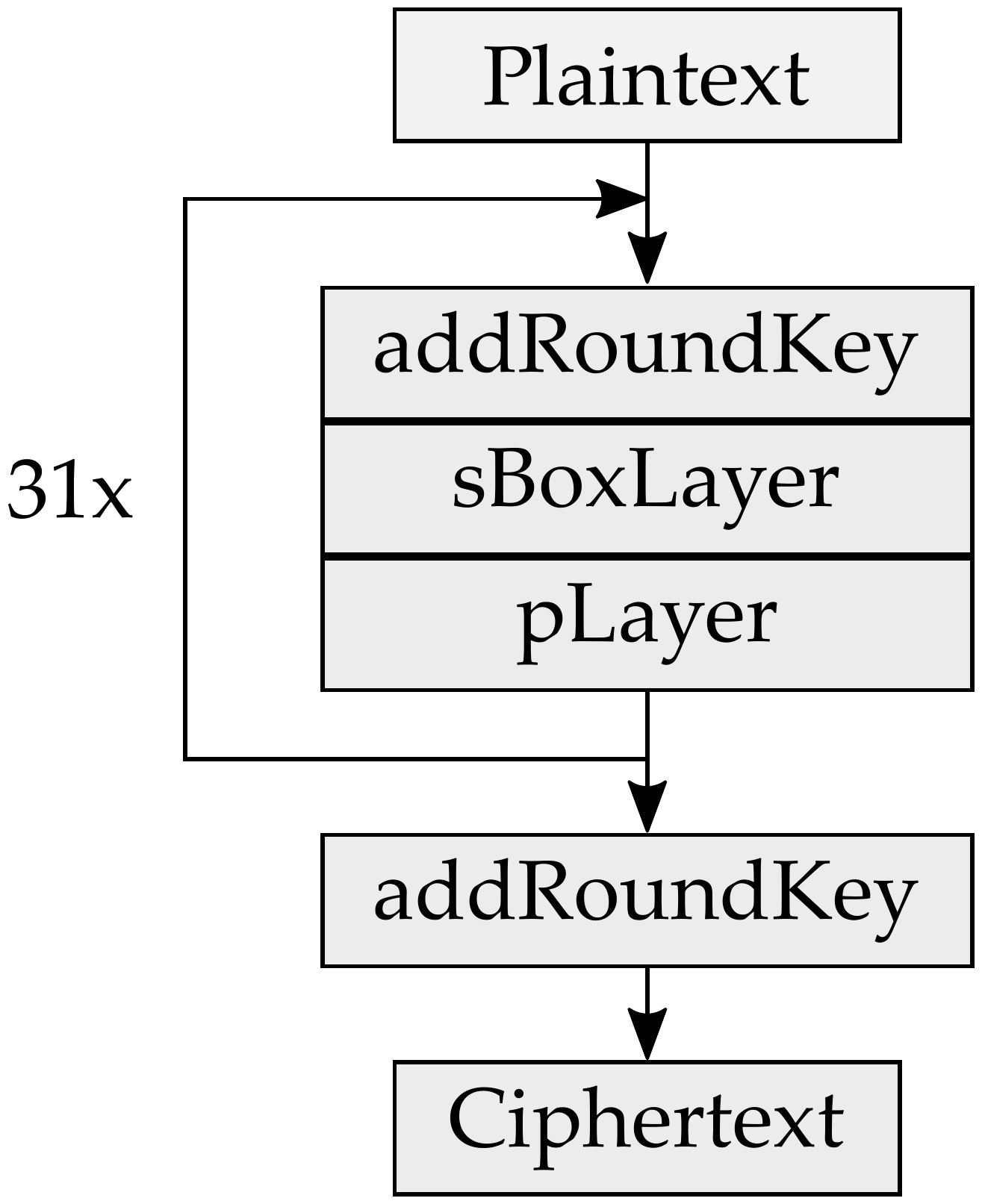}
\caption{High-level overview of operations in PRESENT block cipher.}
\label{fig:present}
\end{figure}

PRESENT is a symmetric block cipher, based on SP-network. It consists of 31
rounds, block length is 64 bits and it supports keys with lengths of 80 and 128
bits. Considering the usage purposes, authors recommend the 80 bit key length
version. Each round consists of three operations: XOR with the round key,
substitution by the 4-bit S-box (Table~\ref{t:sbox}), and bit permutation
(Table~\ref{t:pbox}).
At the end of the cipher, a post-whitening XOR with the key is performed, so there are 32 generated keys
in total. A high-level overview of the encryption process is stated in Figure~\ref{fig:present}.

\begin{table}
\centering
\caption{PRESENT S-box.}
\label{t:sbox}
\ttfamily
\begin{tabular}{|c||c|c|c|c|c|c|c|c|c|c|c|c|c|c|c|c|}
\hline
\emph{x} &0&1&2&3&4&5&6&7&8&9&A&B&C&D&E&F \\ \hline
\emph{S(x)} &C&5&6&B&9&0&A&D&3&E&F&8&4&7&1&2 \\ 
\hline
\end{tabular}
\end{table}

\begin{table}
\centering
\caption{PRESENT permutation layer.}
\label{t:pbox}
\ttfamily
\begin{tabular}{|c||C{0.4cm}|C{0.4cm}|C{0.4cm}|C{0.4cm}|C{0.4cm}|C{0.4cm}|C{0.4cm}|C{0.4cm}|C{0.4cm}|C{0.4cm}|
C{0.4cm}|C{0.4cm}|C{0.4cm}|C{0.4cm}|C{0.4cm}|C{0.4cm}|}
\hline
\emph{i} &0&1&2&3&4&5&6&7&8&9&10&11&12&13&14&15 \\
\emph{P(i)} &0&16&32&48&1&17&33&49&2&18&34&50&3&19&35&51 \\ 
\hline\hline
\end{tabular}
\begin{tabular}{|c||C{0.4cm}|C{0.4cm}|C{0.4cm}|C{0.4cm}|C{0.4cm}|C{0.4cm}|C{0.4cm}|C{0.4cm}|C{0.4cm}|C{0.4cm}|
C{0.4cm}|C{0.4cm}|C{0.4cm}|C{0.4cm}|C{0.4cm}|C{0.4cm}|}
\hline
\emph{i} &16&17&18&19&20&21&22&23&24&25&26&27&28&29&30&31 \\
\emph{P(i)} &4&20&36&52&5&21&37&53&6&22&38&54&7&23&39&55 \\ 
\hline\hline
\end{tabular}
\begin{tabular}{|c||C{0.4cm}|C{0.4cm}|C{0.4cm}|C{0.4cm}|C{0.4cm}|C{0.4cm}|C{0.4cm}|C{0.4cm}|C{0.4cm}|C{0.4cm}|
C{0.4cm}|C{0.4cm}|C{0.4cm}|C{0.4cm}|C{0.4cm}|C{0.4cm}|}
\hline
\emph{i} &32&33&34&35&36&37&38&39&40&41&42&43&44&45&46&47 \\
\emph{P(i)} &8&24&40&56&9&25&41&57&10&26&42&58&11&27&43&59 \\ 
\hline\hline
\end{tabular}
\begin{tabular}{|c||C{0.4cm}|C{0.4cm}|C{0.4cm}|C{0.4cm}|C{0.4cm}|C{0.4cm}|C{0.4cm}|C{0.4cm}|C{0.4cm}|C{0.4cm}|
C{0.4cm}|C{0.4cm}|C{0.4cm}|C{0.4cm}|C{0.4cm}|C{0.4cm}|}
\hline
\emph{i} &48&49&50&51&52&53&54&55&56&57&58&59&60&61&62&63 \\
\emph{P(i)} &12&28&44&60&13&29&45&61&14&30&46&62&15&31&47&63 \\ 
\hline
\end{tabular}
\end{table}

Key schedule of 80-bit version of the algorithm is following. First, a secret
key is stored in register $K$ and represented as $k_{79}k_{78}\ldots k_0$. Since
the block length is 64 bits, only 64 leftmost bits are used for each
round. Therefore, at round $i$ we have:
$$
K_i := \kappa_{63}\kappa_{62}\ldots \kappa_{0} = k_{79}k_{78}\ldots k_{16}.
$$
After every round, key register $K$ is updated in a following way:
\begin{enumerate}[leftmargin=3cm]
  \item $[k_{79}k_{78}\ldots k_1k_0] = [k_{18}k_{17}\ldots k_{20}k_{19}]$
  \item $[k_{79}k_{78}k_{77}k_{76}] = S[k_{79}k_{78}k_{77}k_{76}]$
  \item $[k_{19}k_{18}k_{17}k_{16}k_{15}] =
  [k_{19}k_{18}k_{17}k_{16}k_{15}]\oplus RC$ 
\end{enumerate}
where $RC$ is a round counter.
\section{Attack Model}
\label{attackmodel}
\begin{figure}[t!] \centering
\includegraphics[width=0.9\textwidth]{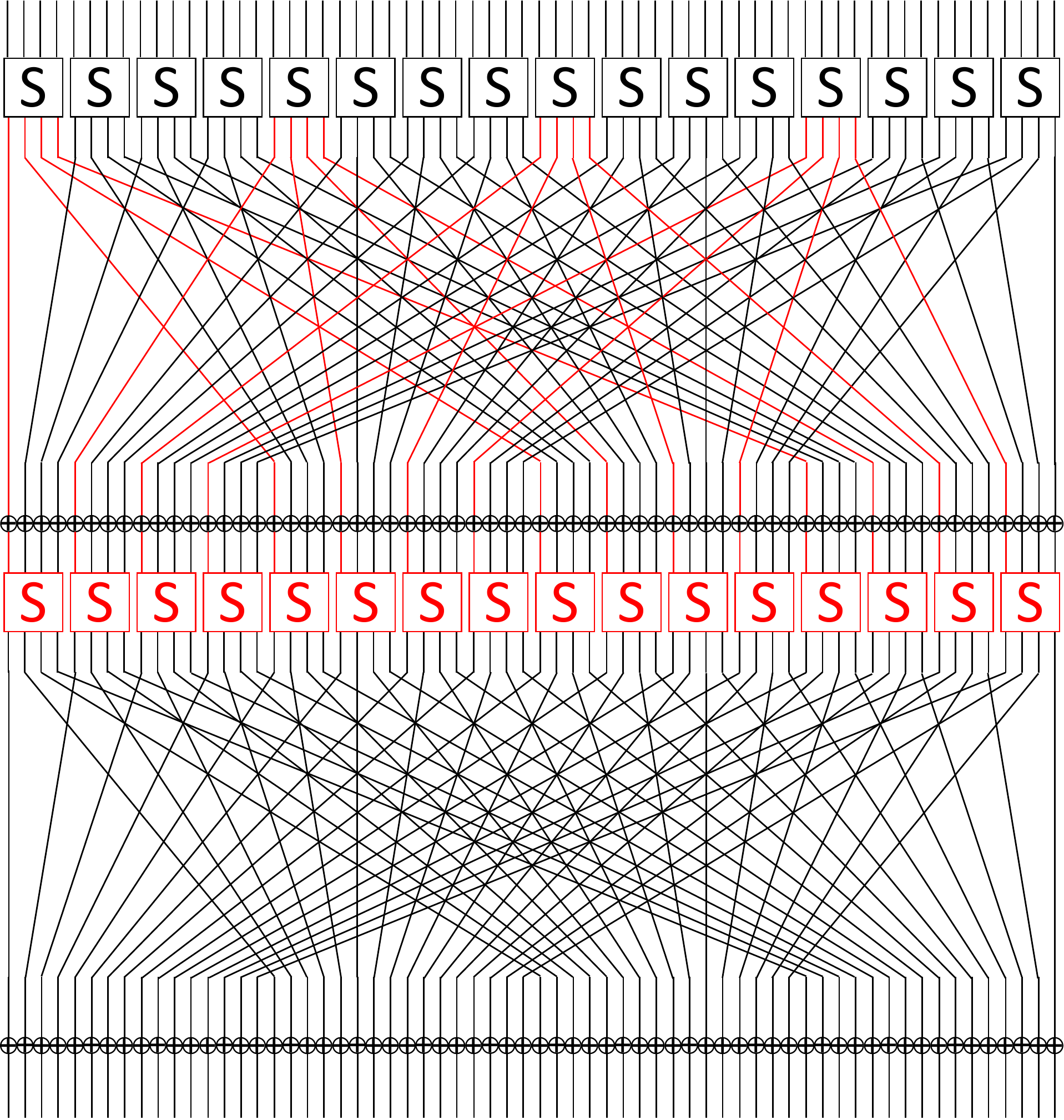}
\caption{Fault propagation after attacking four S-boxes.}
\label{f:present_fault}
\end{figure}

PRESENT algorithm uses sixteen 4-bit S-boxes. The output of S-boxes is an input
for the permutation layer. The attack exploits following properties of the
algorithm:
\begin{itemize}
  \item Output of one S-box is an input for four different S-boxes.
  \item Input of one S-box consists of outputs from four different S-boxes.
  \item There are four different groups of S-boxes: 
  \begin{itemize}
  	\item The outputs of S-boxes 0-3 are inputs for S-boxes 0,4,8,12,
  	\item The outputs of S-boxes 4-7 are inputs for S-boxes 1,5,9,13,
  	\item The outputs of S-boxes 8-11 are inputs for S-boxes 2,6,10,14,
  	\item The outputs of S-boxes 12-15 are inputs for S-boxes 3,7,11,15.
  \end{itemize}
\end{itemize}
If it is possible to corrupt the whole output of some S-box (flip four bits),
the fault will spread into four S-boxes in the following round, affecting the
same bit position in every S-box. Therefore if we aim at four S-boxes from
distinct groups in round 30, the fault will be distributed to every
S-box in round 31. This fact is depicted in Figure.~\ref{f:present_fault}.

\subsection{Attack Steps}
The attack steps are following:
\begin{enumerate}
  \item The attacker inserts four 4-bit faults at the output of four S-boxes
  from distinct groups.
  \item She computes difference tables according to the fault model. More
  specifically, the bit faults in $S_{i,j}^{31}$, where $i\in{0,..,15}$ are
  S-boxes of round 31 and $j\in{0,1,2,3}$ are bits of particular S-box are
  following:
  \begin{itemize}
    \item Fault at $S_0^{30}, S_4^{30}, S_8^{30}, S_{12}^{30}$ will result to
    faults at $S_{i,0}^{31}$.
    \item Fault at $S_1^{30}, S_5^{30}, S_9^{30}, S_{13}^{30}$ will result to
    faults at $S_{i,1}^{31}$.
    \item Fault at $S_2^{30}, S_6^{30}, S_{10}^{30}, S_{14}^{30}$ will result to
    faults at $S_{i,2}^{31}$.
    \item Fault at $S_3^{30}, S_7^{30}, S_{11}^{30}, S_{15}^{30}$ will result to
    faults at $S_{i,3}^{31}$.
  \end{itemize}
  Those difference tables contain the fault mask (bit position of a fault) and
  the output mask, which is the difference of the S-box output of a correct and
  faulty input. As an example, for masks 1000 and 0100, Table
  \ref{t:faults} shows these values.
  \item Using a formula \ref{diff}, the attacker observes the output differences and she
  searches for possible $S^{31}$ input candidates.
  \item She repeats steps 1-3, inserting fault in different nibbles, which will
  result to attacking different S-box bits at $S^{31}$.
  \item Shee compares the possible candidate sets from both attacks, the
  intersection of these sets gives exactly one candidate, which is the correct
  state before $S^{31}$.
  \item After obtaining all the key nibbles of $K^{31}$, the attacker uses an
  exhaustive search on the rest of the key, so the search space is $2^{16}$. 
\end{enumerate}
\begin{equation}
\label{diff}
\Delta = P^{-1}(C') \oplus P^{-1}(C)
\end{equation}

\subsection{Attack Example}
After obtaining the faulty ciphertext and inverting the last round permutation
layer, it is possible to obtain information about the input of the substitution
layer. Let us assume that we attacked the first nibble after the S-box operation
in round 30. Therefore the faulty mask of inputs of S-boxes 0,4,8,12 in round 31
will be 1000. We can now observe the changes in the output of the algorithm.
Table \ref{t:faults} shows every possible input and output of the S-box
operation, together with the faulty one, after attacking the most significant
bit. It is easy to see that in the worst case, we have narrowed the input
candidates to four numbers. For example, for the output difference 1111, we have
input candidates 0, 7, 8, 9. The notation is following:
\begin{itemize}
  \item \textit{I} is a correct input of the S-box.
  \item \textit{I'} is a faulty input of the S-box.
  \item \textit{O} is a correct output of the S-box.
  \item \textit{O'} is a faulty output of the S-box.
  \item $\Delta$ is the output difference between the correct and the faulty
  output.
\end{itemize}

\begin{table}[tb]
\begin{center}
\caption{Faults and differences with fault masks 1000 and 0100.}
\begin{tabular}{ |c| c| c| c| c| } \hline
\multicolumn{5}{c}{Mask 1000} \\ \hline
I & I' & O & O' & $\Delta$ \\ \hline
 0 & 8 & C & 3 & 1111 \\ \hline
 1 & 9 & 5 & E & 1011 \\ \hline
 2 & A & 6 & F & 1001 \\ \hline
 3 & B & B & 8 & 0011 \\ \hline
 4 & C & 9 & 4 & 1101 \\ \hline
 5 & D & 0 & 7 & 0111 \\ \hline
 6 & E & A & 1 & 1011 \\ \hline
 7 & F & D & 2 & 1111 \\ \hline
 8 & 0 & 3 & C & 1111 \\ \hline
 9 & 1 & E & 5 & 1011 \\ \hline
 A & 2 & F & 6 & 1001 \\ \hline
 B & 3 & 8 & B & 0011 \\ \hline
 C & 4 & 4 & 9 & 1101 \\ \hline
 D & 5 & 7 & 0 & 0111 \\ \hline
 E & 6 & 1 & A & 1011 \\ \hline
 F & 7 & 2 & D & 1111 \\ \hline
\end{tabular}
\hspace{0.5cm}
\begin{tabular}{ |c| c| c| c| c| } \hline
\multicolumn{5}{c}{Mask 0100} \\ \hline
I & I' & O & O' & $\Delta$ \\ \hline
 0 & 4 & C & 9 & 0101 \\ \hline
 1 & 5 & 5 & 0 & 0101 \\ \hline
 2 & 6 & 6 & A & 1100 \\ \hline
 3 & 7 & B & D & 0110 \\ \hline
 4 & 0 & 9 & C & 0101 \\ \hline
 5 & 1 & 0 & 5 & 0101 \\ \hline
 6 & 2 & A & 6 & 1100 \\ \hline
 7 & 3 & D & B & 0110 \\ \hline
 8 & C & 3 & 4 & 0111 \\ \hline
 9 & D & E & 7 & 1001 \\ \hline
 A & E & F & 1 & 1110 \\ \hline
 B & F & 8 & 2 & 1010 \\ \hline
 C & 8 & 4 & 3 & 0111 \\ \hline
 D & 9 & 7 & E & 1001 \\ \hline
 E & A & 1 & F & 1110 \\ \hline
 F & B & 2 & 8 & 1010 \\ \hline
\end{tabular}
\label{t:faults}
\end{center}
\end{table}

The second step of the attack is to change the attacked nibble, so that it will
affect another output nibble of the S-box belonging to the same group as the
first one. Let us assume, we attacked the second nibble of the S-box output in
round 30. Therefore the input of the same S-boxes in round 31 will be changed,
but the faulty mask will be 0100 in this case. The outputs for this case are
stated in table \ref{t:faults}. It is easy to see that the groups of the
input values producing the same differences after the fault are not overlapping
with the first attack, therefore we can determine the input value with
certainty. For example if the output difference in this case would be 0101, the
possible input candidates are 0, 1, 4, 5. The only common number for both
attacks is 0, therefore it is the input value for the given S-box.

Using the simplified attack model with only one faulty nibble, it is possible to
reveal the last round key with 8 faulty encryptions, since it is necessary to
attack two distinct nibbles of each of four groups of S-boxes. If there is a
possibility to inject multiple faults per encryption, the last round key could
be revealed with 2 encryptions. In each run, the attacker would inject the fault
into one nibble of the S-box output of four different groups, changing the
nibbles after the first run. In both cases, an exhaustive search of $2^{16}$ is
required to obtain the whole PRESENT-80 key.

\section{Hardware Trojan for Practical Fault Injection}
\label{implementation}
In this section, an FPGA implemented PRESENT cipher for the described multiple 
fault analysis approach will be detailed. The hardware realization relies on 
a Spartan-6 FPGA (XC6SLX4), soldered on Diligent Cmod-S6 commercial board~\cite{digilent2014CmodS6}. We mounted the injections using the hardmacro-based 
Hardware Trojan (HT) \cite{tehranipoor2010survey} by inserting specially created 
trojan modules into the algorithmic networks in FPGA scenario.
\subsection{FPGA Scenario}
\begin{figure}[!h]  
\centering
\includegraphics[width=3.5in]{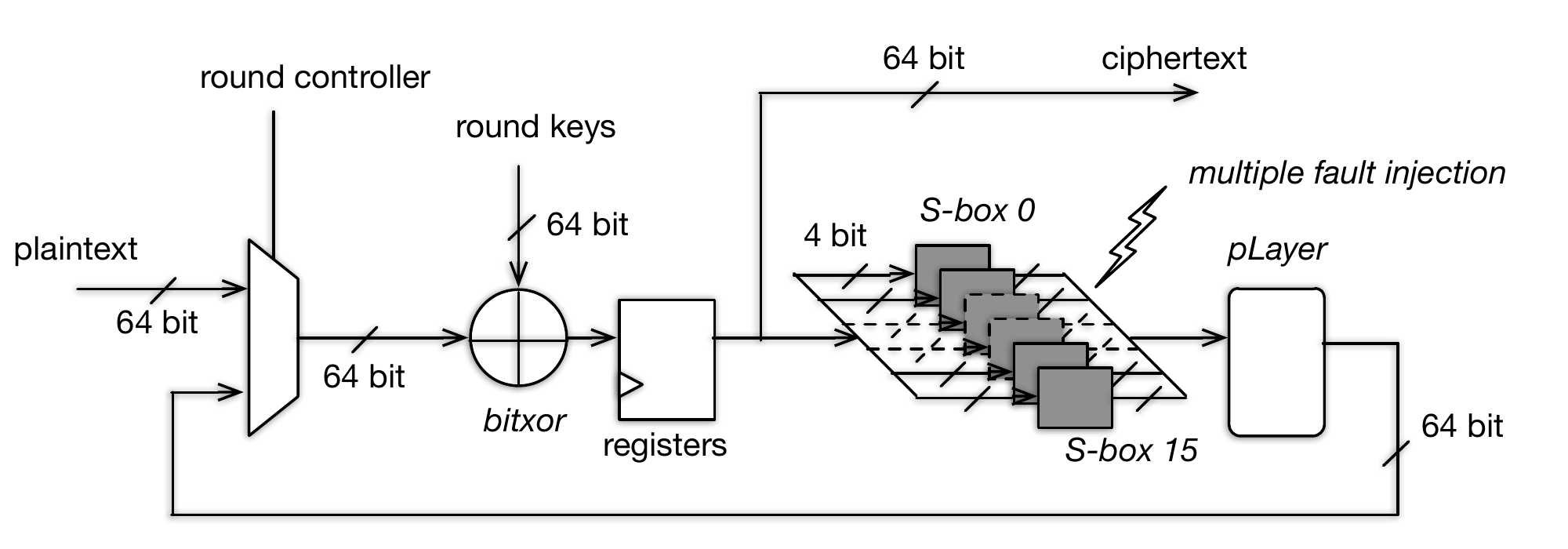}
\caption{Parallel \emph{Present} block cipher in FPGA.}
\label{1_present_architecture}
\end{figure}

Field Programmed Gate Array (FPGA) has been widely utilized in almost 
all digital/hybrid logic applications due to its rapid implementation, 
low cost and high performance. The major advantage of hardware implementation 
for cipher lies within its parallel computation networks that allow multiple 
logic chains computed in parallel, which results in high computational speed 
compared to the microcontroller scenarios. In our work, PRESENT cipher 
was implemented inside a compact Spartan-6 FPGA. 
The cipher is structured in loops with 16 4-bit S-boxes in parallel. The 
complete encryption is clocked with a global clock signal and the intermediate 
values from each round are stored in 128 1-bit registers. So a complete encryption in out implementation consists of 32 clock cycles, as seen in Figure~\ref{1_present_architecture}.

\subsection{Hardware Trojan for Fault Injection}
\begin{figure}[t!]  
\centering
\includegraphics[width=1.0\textwidth]{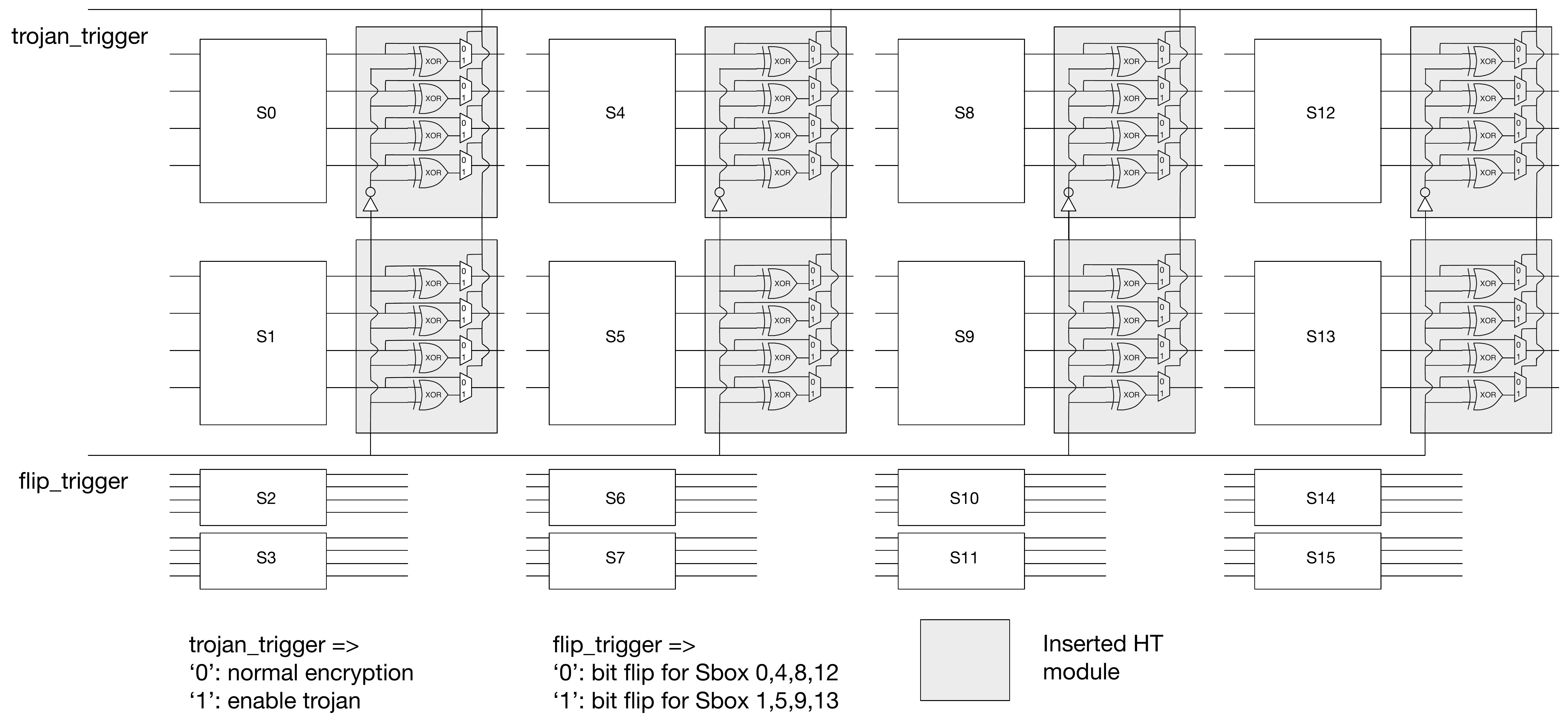}
\caption{Trojans inserted into the signal paths after specific S-boxes.}
\label{2_trojan}
\end{figure}

HT typically requires some trigger signals to activate the inserted trojan 
modules. By the principle of the proposed multiple fault injection, a 1 bit 
signal of \emph{trojan\_trigger} is activated at the 30th encryption 
round when the fault perturbation is required. Since the injection point 
is at the output of the S-box, the specially devised trojan hardmacro for 
S-boxes 0, 4, 8, 12, and for S-boxes 1, 5, 9, 13 can be inserted during the 
chip fabrication or at the off-the-shelf stage. The payload is highlighted 
in the grey box of Figure~\ref{2_trojan}. Signal of the \emph{flip\_trigger} 
is used to control the injection into different S-box groups. Since only 2 
injection rounds are needed for the proposed fault attack, another 1 bit 
signal is enough. So there are totally 2 bit signals acting as the trigger 
in this solution. In Spartan-6 FPGA, every slice consists of 4 look-up tables 
(LUT), 8 multiplexers and 8 flip-flops. Even each LUT can either be used as one 
6-input 1-output Boolean function or two 5-input 1-output functions, only 4 multiplexers in each slice has an external input that we have to use as the 1 input bit of the XOR gate and the input of multiplexer after the XOR gate. Therefore
one slice can actually implement 4 HT modules, and 8 extra slices, \emph{i.e.,} 4 
Configurable Logic Blocks (CLBs), are sufficient for inserting all the 8 trojans. 
The states of the 2-bit trigger signals are given in Table~\ref{HT_trigger_state}. 
It is emphasized that the insertion of trojan can as well apply to different 
S-box groups, just obeying the principles explained in Section~\ref{attackmodel}.

\begin{table}
\centering
\caption{Trojan Trigger State.}
\begin{tabular}{ p{1.5cm}<\centering p{0.8cm}<\centering| p{2cm}<\centering|
p{2.5cm}<\centering} 
& &  \multicolumn{2}{c}{trojan\_trigger} \\ \cline{3-4}
& & `0' & `1' \\
\cline{1-4}
\multirow{2}{*}{flip\_trigger}
&\multicolumn{1}{|c|}{`0'}&\multirow{2}{*}{encryption}&1st insertion
\\
&\multicolumn{1}{|c|}{`1'}&&2nd insertion\\
\end{tabular}
\label{HT_trigger_state}
\end{table}

\section{Conclusions}
\label{conclusions}
In our paper we have proposed a multiple fault injection attack on PRESENT cipher, using DFA technique. By flipping four nibbles in the penultimate round
we were able to obtain the secret key using two faulty ciphertext and an exhaustive search of a $2^{16}$ bits. We have implemented a hardware trojan,
causing this type of fault on an FPGA implementation of PRESENT.

There are two other possible attack scenarios. The first one is flipping more
than 4 bits, therefore the attack would affect the following nibble as well. In
this case, more than one bit of particular S-box in round 31 will be affected.
The attack can be still executed in the same way, only the differential table
for the concrete fault mask has to be computed. It can be shown, that
intersection of arbitrary two masks from two distinct differential tables gives
exactly one candidate in every case, so it is possible to use multiple-bit fault
masks.

The other scenario is flipping less than 4 bits in one nibble. In this case, the
fault doesn't spread into every $S^{31}$ S-box. The solution for this scenario
is to use more faults, therefore the number of encryptions will increase.

The hardware trojan approach supporting the proposed multiple fault attack relies on the hardmacro based HT modules to be inserted into the outputs of 
specific S-boxes. In our tested Spartan-6 FPGA, only 4 CLBs are 
sufficient to implement the 8 trojan modules combined with 2 bit global trigger signal. 
Since the trojan blocks are inserted into the signal path of specific S-box outputs, without altering the main functionality of the ciphers, the trojan can be mounted at almost all design stages, such as front-end HDL coding or netlist alteration, as well as back-end layout or sub-gate manipulation.
%

In the subsequent work, we would like to focus on the realistic trojan insertion using the proposed multiple fault analysis into the security modules of an IoT scenario, and the related HT detection technique will also be emphasized. 

\bibliographystyle{splncs03}
\bibliography{sigproc} 

\begin{thebibliography}{10}
\providecommand{\url}[1]{\texttt{#1}}
\providecommand{\urlprefix}{URL }

\bibitem{present_dfa3}
Bagheri, N., Ebrahimpour, R., Ghaedi, N.: {New differential fault analysis on
  PRESENT}. EURASIP Journal on Advances in Signal Processing  2013(1),  1--10
  (2013)

\bibitem{biham_shamir}
Biham, E., Shamir, A.: Differential fault analysis of secret key cryptosystems.
  In: Kaliski, BurtonS., J. (ed.) Advances in Cryptology — CRYPTO '97, Lecture
  Notes in Computer Science, vol. 1294, pp. 513--525. Springer Berlin
  Heidelberg (1997)

\bibitem{blondeau2014present}
Blondeau, C., Nyberg, K.: Links between truncated differential and
  multidimensional linear properties of block ciphers and underlying attack
  complexities. In: Oswald, E., Nguyen, P.Q. (eds.) Eurocrypt 2014. Lecture
  Notes in Computer Science, vol. 8441. Springer-Verlag (2014)

\bibitem{present}
Bogdanov, A., Knudsen, L.R., Leander, G., Paar, C., Poschmann, A., Robshaw,
  M.J., Seurin, Y., Vikkelsoe, C.: {PRESENT: An Ultra-Lightweight Block
  Cipher}. In: Proceedings of the 9th International Workshop on Cryptographic
  Hardware and Embedded Systems. pp. 450--466. CHES '07, Springer-Verlag,
  Berlin, Heidelberg (2007)

\bibitem{boneh_demillo}
Boneh, D., DeMillo, R.A., Lipton, R.J.: On the importance of checking
  cryptographic protocols for faults. In: Proceedings of the 16th Annual
  International Conference on Theory and Application of Cryptographic
  Techniques. pp. 37--51. EUROCRYPT'97, Springer-Verlag, Berlin, Heidelberg
  (1997)

\bibitem{chakraborty2013hardware}
Chakraborty, R.S., Saha, I., Palchaudhuri, A., Naik, G.K.: Hardware trojan
  insertion by direct modification of fpga configuration bitstream. Design \&
  Test, IEEE  30(2),  45--54 (2013)

\bibitem{clavier2012}
Clavier, C.: {Attacking Block Ciphers}. In: Joye, M., Tunstall, M. (eds.) Fault
  Analysis in Cryptography, pp. 19--35. Information Security and Cryptography,
  Springer Berlin Heidelberg (2012)

\bibitem{danger2013}
Danger, J.L., Guilley, S., Hoogvorst, P., Murdica, C., Naccache, D.: A
  synthesis of side-channel attacks on elliptic curve cryptography in
  smart-cards. Journal of Cryptographic Engineering  3(4),  241--265 (2013)

\bibitem{defense2005defense}
Defense, U.O.: Defense science board task force on high performance microchip
  supply. Washington, DC pp. 2005--02 (2005)

\bibitem{digilent2014CmodS6}
Digilent: Cmod s6 reference manual, xilinx. Inc., October  (Jan 30, 2014)

\bibitem{dinu2015lightweight}
Dinu, D., Le~Corre, Y., Khovratovich, D., Perrion, L., Gro{\ss}sch{\"a}dl, J.,
  Biryukov, A.: Triathlon of lightweight block ciphers for the internet of
  things. Cryptology ePrint Archive, Report 2015/209 (2015),
  http://eprint.iacr.org/2015/209.pdf

\bibitem{present_dfa4}
Gu, D., Li, J., Li, S., Ma, Z., Guo, Z., Liu, J.: Differential fault analysis
  on lightweight blockciphers with statistical cryptanalysis techniques. In:
  Fault Diagnosis and Tolerance in Cryptography (FDTC), 2012 Workshop on. pp.
  27--33 (Sept 2012)

\bibitem{present_dfa5}
Jeong, K., Lee, Y., Sung, J., Hong, S.: Improved differential fault analysis on
  present-80/128. International Journal of Computer Mathematics  90(12),
  2553--2563 (2013)

\bibitem{jin2009experiences}
Jin, Y., Kupp, N., Makris, Y.: Experiences in hardware trojan design and
  implementation. In: Hardware-Oriented Security and Trust, 2009. HOST'09. IEEE
  International Workshop on. pp. 50--57. IEEE (2009)

\bibitem{katagi2008iot}
Katagi, M., Moriai, S.: Lightweight cryptography for the internet of things.
  Sony Corporation  (2008)

\bibitem{king2008designing}
King, S.T., Tucek, J., Cozzie, A., Grier, C., Jiang, W., Zhou, Y.: Designing
  and implementing malicious hardware. LEET  8,  1--8 (2008)

\bibitem{lin2009trojan}
Lin, L., Kasper, M., G{\"u}neysu, T., Paar, C., Burleson, W.: Trojan
  side-channels: Lightweight hardware trojans through side-channel engineering.
  In: Cryptographic Hardware and Embedded Systems-CHES 2009, pp. 382--395.
  Springer (2009)

\bibitem{shiyanovskii2010process}
Shiyanovskii, Y., Wolff, F., Rajendran, A., Papachristou, C., Weyer, D., Clay,
  W.: Process reliability based trojans through nbti and hci effects. In:
  Adaptive Hardware and Systems (AHS), 2010 NASA/ESA Conference on. pp.
  215--222. IEEE (2010)

\bibitem{tehranipoor2010survey}
Tehranipoor, M., Koushanfar, F.: A survey of hardware trojan taxonomy and
  detection  (2010)

\bibitem{tehranipoor2011introduction}
Tehranipoor, M., Wang, C.: Introduction to hardware security and trust.
  Springer Science \& Business Media (2011)

\bibitem{torrance2011state}
Torrance, R., James, D.: The state-of-the-art in semiconductor reverse
  engineering. In: Proceedings of the 48th Design Automation Conference. pp.
  333--338. ACM (2011)

\bibitem{present_dfa1}
Wang, G., Wang, S.: {Differential Fault Analysis on PRESENT Key Schedule}. In:
  Computational Intelligence and Security (CIS), 2010 International Conference
  on. pp. 362--366 (Dec 2010)

\bibitem{present_dfa2}
Zhao, X., Guo, S., Wang, T., Zhang, F., Shi, Z.: {Fault-propagate pattern based
  DFA on PRESENT and PRINT cipher}. Wuhan University Journal of Natural
  Sciences  17(6),  485--493 (2012)

\end{thebibliography}

\end{document}